\documentstyle[12pt,epsfig]{article} 

\newcommand{\agt}{\,\rlap{\lower 3.5 pt \hbox{$\mathchar \sim$}} \raise 1pt
 \hbox {$>$}\,}
\newcommand{\alt}{\,\rlap{\lower 3.5 pt \hbox{$\mathchar \sim$}} \raise 1pt
 \hbox {$<$}\,}

\catcode`@=11
\newcount\@tempcntc
\def\@citex[#1]#2{\if@filesw\immediate\write\@auxout{\string\citation{#2}}\fi
  \@tempcnta\z@\@tempcntb\m@ne\def\@citea{}\@cite{\@for\@citeb:=#2\do
    {\@ifundefined
       {b@\@citeb}{\@citeo\@tempcntb\m@ne\@citea\def\@citea{,}{\bf ?}\@warning
       {Citation `\@citeb' on page \thepage \space undefined}}%
    {\setbox\z@\hbox{\global\@tempcntc0\csname b@\@citeb\endcsname\relax}%
     \ifnum\@tempcntc=\z@ \@citeo\@tempcntb\m@ne
       \@citea\def\@citea{,}\hbox{\csname b@\@citeb\endcsname}%
     \else
      \advance\@tempcntb\@ne
      \ifnum\@tempcntb=\@tempcntc
      \else\advance\@tempcntb\m@ne\@citeo
      \@tempcnta\@tempcntc\@tempcntb\@tempcntc\fi\fi}}\@citeo}{#1}}
\def\@citeo{\ifnum\@tempcnta>\@tempcntb\else\@citea\def\@citea{,}%
  \ifnum\@tempcnta=\@tempcntb\the\@tempcnta\else
   {\advance\@tempcnta\@ne\ifnum\@tempcnta=\@tempcntb \else \def\@citea{--}\fi
    \advance\@tempcnta\m@ne\the\@tempcnta\@citea\the\@tempcntb}\fi\fi}
\catcode`@=12

\begin{document}

\title{\vskip-3cm{\baselineskip14pt
\centerline{\normalsize DESY 01-071\hfill ISSN 0418-9833}
\centerline{\normalsize KA-TP-21-2001\hfill}
\centerline{\normalsize hep-ph/0109110\hfill}
\centerline{\normalsize May 2001\hfill}
}
\vskip1.5cm
On-Mass-Shell Renormalization of Fermion Mixing Matrices}
\author{K.-P. O. Diener$^a$, B. A. Kniehl$^b$\\
{\normalsize $^a$ Institut f\"ur Theoretische Physik,
Universit\"at Karlsruhe,}\\
{\normalsize Engesserstra\ss e 7, 76131 Karlsruhe, Germany}\\
{\normalsize $^b$ II. Institut f\"ur Theoretische Physik,
Universit\"at Hamburg,}\\
{\normalsize Luruper Chaussee 149, 22761 Hamburg, Germany}}

\date{}

\maketitle

\thispagestyle{empty}

\begin{abstract}
We consider favourable extensions of the standard model (SM) where the lepton
sector contains Majorana neutrinos with vanishing left-handed mass terms, thus
allowing for the see-saw mechanism to operate, and propose physical
on-mass-shell (OS) renormalization conditions for the lepton mixing matrices
that comply with ultraviolet finiteness, gauge-parameter independence, and
(pseudo)unitarity.
A crucial feature is that the texture zero in the neutrino mass matrix is
preserved by renormalization, which is not automatically the case for
possible generalizations of existing renormalization prescriptions for the 
Cabibbo-Kobayashi-Maskawa (CKM) quark mixing matrix in the SM.
Our renormalization prescription also applies to the special case of the SM
and leads to a physical OS definition of the renormalized CKM matrix.
\medskip

\noindent
PACS numbers: 11.10.Gh, 13.38.Be, 14.60.Pq, 14.60.St
\end{abstract}

\newpage

\section{Introduction}\label{sec:one}

The Cabibbo-Kobayashi-Maskawa (CKM) \cite{cab} mixing matrix, which rules the
charged-current interactions of the quark mass eigenstates and enables the 
heavier ones to decay to the lighter ones, is one of the central
ingredients of the standard model (SM) of elementary particle physics and, in
particular, it is the key to our understanding why certain laws of nature are
not invariant under simultaneous charge-conjugation and parity
transformations.
The CKM matrix is customarily parameterized by three angles and one phase,
which constitute four basic parameters of the SM and must be determined by 
experiment.
These parameters represent constants of nature and are listed in the Review of
Particle Physics \cite{pdg}.
The CKM matrix elements appear in the bare SM Lagrangian and are thus subject
to renormalization.
This was realized for the Cabibbo angle in the SM with two fermion generations
in a pioneering paper by Marciano and Sirlin \cite{mar} and for the CKM matrix
of the three-generation SM by Denner and Sack \cite{den} more than a decade 
ago.
So far, all experimental determinations of CKM matrix elements are based on
formulas that do not take this into account \cite{pdg,hoe}.

In quantum electrodynamics, it is very natural and convenient to choose the
on-mass-shell (OS) renormalization scheme, which uses the fine-structure
constant measured in Thomson scattering and the pole masses of the physical
particles as basic parameters.
When one attempts to generalize this renormalization scheme to the SM, one
also needs to specify a suitable, physically motivated renormalization
condition for the CKM matrix.
What are the desired properties of the latter?
As usual, we split the bare CKM matrix elements $V_{ij}^0$, which appear in
the original SM Lagrangian, into their renormalized counterparts $V_{ij}$ and
the counterterms $\delta V_{ij}$ as $V_{ij}^0=V_{ij}+\delta V_{ij}$.
Here, $i$ and $j$ label the generations of the up-type and down-type quarks,
respectively.
To start with, we remark that the parameters $V_{ij}^0$ are ultraviolet (UV)
divergent beyond the tree level and gauge independent.
Furthermore, they form a unitary matrix,
\begin{equation}
\sum_kV_{ik}^0V_{kj}^{0\dagger}=\sum_kV_{ik}^{0\dagger}V_{kj}^0=\delta_{ij},
\end{equation}
which follows from their very definition,
$V_{ij}^0=\sum_kU_{ik}^{0,u}U_{kj}^{0,d\dagger}$, in terms of the unitary
matrices $U^{0,u}$ and $U^{0,d}$ that rotate the weak-interaction eigenstates
of the bare left-handed up-type and down-type quark fields, respectively, into
their mass eigenstates.
The UV divergences of $V_{ij}^0$ are {\it a priori} unknown.
\begin{enumerate}
\item
Clearly, $\delta V_{ij}$ must cancel the UV divergences that, upon 
coupling and mass renormalization, are left in the loop-corrected amplitude of
an arbitrary physical process involving quark mixing.
This requirement fixes the UV divergences of $\delta V_{ij}$.
Different renormalization schemes then differ in the finite parts of
$\delta V_{ij}$.
\item
Apart from being finite, the parameters $V_{ij}$ should also be gauge
independent, so that they qualify as proper physical observables that can be
extracted from experiment with reason.
There is a yet more fundamental reason for this requirement.
In fact, for one physical process, namely the decay $W^+\to u_i\bar d_j$,
where $u_i$ and $d_j$ denote generic up-type and down-type quarks,
respectively, it was shown by explicit calculation to one loop in the OS
renormalization scheme adopting the $R_\xi$ gauge \cite{fuj} that the
loop-corrected transition ($T$) matrix element is gauge independent (but UV
divergent) if all counterterms are included, except for $\delta V_{ij}$
\cite{mad,gam,kni,bar}.
Consequently, the loop-corrected result for the partial width of the decay
$W^+\to u_i\bar d_j$ would be gauge dependent if $\delta V_{ij}$ were.
However, this must not be the case.
\item
On general grounds, renormalization should be arranged so that the basic
structure of the theory is preserved.
Since the bare CKM matrix is unitary, the same should, therefore, be true for
its renormalized version.
Otherwise, four real input parameters would not be sufficient to parameterize
the latter, and the familiar notion of unitary triangle would discontinue to
be meaningful beyond the tree level.
In turn, this would jeopardize the Becchi-Rouet-Stora \cite{brs} symmetry of
the theory \cite{kra}.
The unitarity of the renormalized CKM matrix also follows from the request 
that the commutation relations of the local functional operators related to
the Ward-Takahashi \cite{wt} identities of the theory formulated with
background fields be preserved \cite{gam}.
At one loop, this leads us to require that
\begin{equation}
\sum_k\left(\delta V_{ik}V_{kj}^\dagger+V_{ik}\delta V_{kj}^\dagger\right)=
\sum_k\left(\delta V_{ik}^\dagger V_{kj}+V_{ik}^\dagger\delta V_{kj}\right)=0
\label{eq:uni}
\end{equation}
is valid up to higher-order terms.
\end{enumerate}
In summary, a reasonable OS renormalization prescription for the CKM matrix
should be physically motivated and satisfy the three requirements enumerated
above: UV finiteness, gauge independence, and unitarity.
For aesthetical reasons, we wish to add the optional requirement that all
pairs $(i,j)$ be treated on a democratic footing.

The OS renormalization prescription for the CKM matrix proposed in 
Ref.~\cite{den} is compact and plausible, complies with the first and third
criteria by construction, but --- at first sight surprisingly --- it fails to
satisfy the second criterion because the finite terms of the proposed
expressions for $\delta V_{ij}$ are gauge dependent, as was noticed only
recently \cite{mad,gam,kni,bar}.
Obviously, this problem can be circumvented by adopting the modified
minimal-subtraction ($\overline{\rm MS}$) renormalization scheme \cite{bur} in
dimensional regularization \cite{bol}, where one only retains the UV
divergences of $\delta V_{ij}$, proportional to $2/(4-D)+\ln(4\pi)-\gamma_E$.
Here, $D$ is the space-time dimensionality, and $\gamma_E$ is Euler's
constant.
In Refs.~\cite{mad,gam}, an alternative, OS-like prescription was proposed
that avoids this problem at one loop.
The characteristic feature of this prescription is that the quark
self-energies that enter the definitions of $\delta V_{ij}$ are not evaluated
on their respective mass shells, but at the common subtraction point $q^2=0$.
In Ref.~\cite{kni}, this prescription was adopted to calculate the partial
decay widths of the $W$ boson at one loop in the OS renormalization scheme.
Recently, two further prescriptions were introduced \cite{bar,yam}.
The prescription of Ref.~\cite{bar} is formulated with reference to the case
of zero mixing.
As will be demonstrated in Section~\ref{sec:two}, it does not comply with the
third criterion.
However, we will explain how this drawback can be eliminated.
In Ref.~\cite{yam}, the prescription of Ref.~\cite{den} was modified by
rearranging the off-diagonal quark wave-function renormalization constants in
a manner similar to the pinch technique so that the second criterion is 
satisfied.

Although the renormalization of the CKM matrix is relevant from the conceptual
point of view, its phenomenological significance is damped by the fact that 
the resulting one-loop effects are at most of order
$(\alpha/\pi)m_b^2/M_W^2\approx10^{-5}$, where $\alpha$ is the fine-structure 
constant and $m_b$ and $M_W$ are the masses of the bottom quark and the $W$
boson, respectively \cite{den,kni}.
This may be understood by observing that, in the approximation of neglecting
the masses of the down-type quarks against the $W$-boson mass, the CKM matrix
can be taken to be unity, so that it does not need to be renormalized at all.
The situation is possibly very different for lepton mixing in a non-minimal SM
with massive Dirac neutrinos or in extensions of the SM involving Majorana
neutrinos.
We recall that the three neutrino flavours of the minimal SM are strictly
massless by construction, due to the absence of right-handed neutrino states.

Recent experiments with solar and atmospheric neutrinos \cite{kaj} suggest that 
the known three neutrinos have nonvanishing, yet very small masses and
oscillate.
In fact, there are experimental indications \cite{kaj} that mixing is maximal.
An attractive theoretical framework for such a scenario is provided by the
see-saw mechanism \cite{yan}, which requires the existence of right-handed
neutrino states in addition to the left-handed ones of the minimal SM.
A suitably extended SM Lagrangian contains very large right-handed 
Majorana-neutrino masses, which, together with Dirac-neutrino masses of the
order of the charged-lepton or quark masses, form the non-vanishing entries of
the see-saw mass matrix.
On the other hand, left-handed Majorana-neutrino mass terms are absent, since 
they would break the gauge invariance of the theory \cite{sch}.
This leads to the typical texture zero in the see-saw mass matrix.
Diagonalization of the latter then naturally gives rise to non-zero, but very
small masses for the three known neutrinos, in agreement with experiment, as
well as to ultra-heavy neutrinos, which have not yet been discovered.
The Lagrangian of this class of Majorana-neutrino theories may be found in
Ref.~\cite{apo}; see also Ref.~\cite{pil}.
Since lepton mixing effects are essential in such extensions of the SM, it
is indispensable to renormalize the lepton mixing matrices, not only from the
conceptual point of view, but also from the phenomenological one.

In Ref.~\cite{pil}, the OS renormalization prescription of Ref.~\cite{den} was
extended to general theories with interfamily mixing of Dirac and/or Majorana
fermions.
By construction, this extended prescription complies with the first and third
criteria.
However, we found \cite{die} that the failure of the prescription of
Ref.~\cite{den} to satisfy the second criterion carries over to the one of
Ref.~\cite{pil}.
Of course, this problem could be avoided by adopting the $\overline{\rm MS}$
renormalization scheme \cite{bur}.

In this paper, we propose a novel OS renormalization prescription for the
lepton mixing matrices of Majorana-neutrino theories that is physically
motivated and complies with all three criteria \cite{die}.
We concentrate on the see-saw scenario described above, which is arguably most
appropriate to describe the present experimental situation.
A crucial feature of our prescription is that the texture zero in the neutrino
mass matrix is preserved by renormalization.
We stress that this additional condition is in general not fulfilled for
possible generalizations of existing renormalization prescriptions for the CKM
matrix.
Our prescription also applies to the special case of the SM and leads to a new
OS definition of the renormalized CKM matrix.

This paper is organized as follows.
In Section~\ref{sec:two}, we develop our new OS renormalization prescription
for the special case of the CKM matrix in the SM.
In Section~\ref{sec:three}, we generalize it to the Majorana-neutrino theories
based on the see-saw mechanism described above.
Section~\ref{sec:four} contains our conclusions.

\section{Renormalization of the CKM matrix in the SM}\label{sec:two}

In this section, we reconsider the partial width of the decay
$W^+\to u_i\bar d_j$ at one loop in the SM and develop a new OS
renormalization prescription for the CKM matrix.
We adopt the notation from Ref.~\cite{kni}.
The one-loop-corrected $T$-matrix element has the following structure
\cite{kni}:\footnote{%
In Ref.~\cite{kni}, the expression in the second line of Eq.~(2.6) should be
multiplied by the overall factor $1/V_{ij}$.}
\begin{eqnarray}
{\cal M}_1^{Wu_id_j}&=&-\frac{eV_{ij}}{\sqrt2s_w}\left\{
{\cal M}_1^-\left[1+\frac{\delta e}{e}-\frac{\delta s_w}{s_w}
+\frac{\delta V_{ij}}{V_{ij}}+\frac{1}{2}\delta Z_W\right.\right.
\nonumber\\
&&{}+\left.\left.
\frac{1}{2V_{ij}}\sum_k\left(\delta Z_{ik}^{u,L\dagger}V_{kj}
+V_{ik}\delta Z_{kj}^{d,L}\right)\right]
+\sum_{a=1}^2\sum_{\sigma=\pm}{\cal M}_a^\sigma
\delta F_a^\sigma\left(M_W,m_{u,i},m_{d,j}\right)\right\},\quad
\label{eq:amp}
\end{eqnarray}
where $e=\sqrt{4\pi\alpha}$ is the electron charge magnitude,
$s_w=\sin\theta_w$ is the sine of the weak mixing angle, $\delta Z_W$ is the
$W$-boson wave-function renormalization constant, $\delta Z_{ij}^{u,L}$ and
$\delta Z_{ij}^{d,L}$ are the left-handed wave-function renormalization
constants for the up-type and down-type quarks, respectively,
${\cal M}_a^\sigma$ are standard matrix elements expressed in terms of the
four-momenta, polarization four-vectors, and spinors of the $W$ boson and the
$u_i$ and $\bar d_j$ quarks, and $\delta F_a^\sigma$ are electroweak form
factors arising from the proper vertex corrections.
The label $\sigma=\pm$ refers to right-handed/left-handed chirality.
Notice that the proper vertex corrections only depend linearly on $V_{ij}$,
which is factored out in Eq.~(\ref{eq:amp}), because, due to electric-charge
conservation, there is just one $W^+u_i\bar d_j$ vertex in each one-loop
triangle diagram.
We stress that the linear dependence of the vertex corrections on the fermion
mixing matrix is a special feature of the SM at one loop, which ceases to be
true at higher orders in the SM or even at one loop in general extensions of
the SM involving Majorana neutrinos.
The renormalization constants $\delta e$, $\delta s_w$, $\delta Z_W$,
$\delta Z_{ij}^{u,L}$, and $\delta Z_{ij}^{d,L}$ are all uniquely defined in
the electroweak OS renormalization scheme \cite{sir}.
In the following, we determine $\delta V_{ij}$ from a physical OS
renormalization condition so that all three criteria enumerated in
Section~\ref{sec:one} are satisfied.

In order to fix $\delta V_{ij}$, we proceed in two steps.
In the first step, we impose a physical OS renormalization condition to
construct an intermediate expression $\delta\tilde V_{ij}$ that contains the
correct UV divergences and is gauge independent.
In other words, we split $V_{ij}^0=\tilde V_{ij}+\delta\tilde V_{ij}$ in such 
a way that $\tilde V_{ij}$ satisfies the first and second criteria, but not
necessarily the third one.
In the second step, we shift $\delta\tilde V_{ij}$ by UV-finite,
gauge-independent terms, so that also the third criterion is satisfied.

Our formalism to fix $\delta\tilde V_{ij}$ is based on the simple observation
that, in the presence of quark mixing, the mapping of up-type and down-type
quark mass eigenstates into doublets is completely arbitrary.
In the standard nomenclature, the quark mass eigenstates are associated with
fermion generations in the order of their masses.
However, this is but a convention, albeit a reasonable one.
Our proposal is to define $\delta\tilde V_{ij}$ by matching Eq.~(\ref{eq:amp})
with its counterpart in the theory that emerges from the SM by turning off
quark mixing and treating $u_i$ and $d_j$ as isopartners.
To be specific, the bare Lagrangian of this modified theory emerges from the
one of the SM by taking the bare CKM matrix to be the unit matrix and 
interchanging the down-type quark fields $d_i$ and $d_j$ if $i\ne j$.
This is equivalent to substituting in the bare SM Lagrangian the expression
\begin{equation}
V_{kl}^0=\delta_{ij}\delta_{kl}+(1-\delta_{ij})(\delta_{ik}\delta_{jl}
+\delta_{il}\delta_{jk}+\epsilon_{ijk}\epsilon_{ijl}),
\label{eq:v0}
\end{equation}
where it is understood that indices that appear in a product more than once 
are not summed over.
Notice that $i$ and $j$ are singled out in this modified theory.
Each particular choice of $i$ and $j$ defines a different such theory.
Since the bare CKM matrix of Eq.~(\ref{eq:v0}) only contains the entries zero 
and one, it does not need to be renormalized.
Thus, the OS renormalization of the modified theory is uniquely fixed by the
well-established procedure \cite{sir}.
In particular, the renormalized electron charge magnitude and particle masses
are identified with the respective constants of nature.
In turn, this implies that they coincide with those of the SM.
On the other hand, the renormalization constants of the parameters and fields
of the modified theory will in general differ from their counterparts in the
SM because the Feynman rules for the $W^+u_k\bar d_l$ and $W^-\bar u_kd_l$
vertices are different.
As for the parameters, this must be compensated by appropriate shifts in the
bare quantities.
Henceforth, we denote the renormalization constants and bare parameters of the
modified theory by a caret.
In particular, we have $e=e^0-\delta e=\hat e^0-\delta\hat e$ and
$s_w=s_w^0-\delta s_w=\hat s_w^0-\delta\hat s_w$.\footnote{%
At one loop, we have $\delta\hat e=\delta e$ and $\hat e^0=e^0$, since the
CKM matrix does not yet enter the electric-charge renormalization.}
In the reference theory, Eq.~(\ref{eq:amp}) is thus replaced by
\begin{eqnarray}
\hat{\cal M}_1^{Wu_id_j}&=&-\frac{e}{\sqrt2s_w}\left\{
{\cal M}_1^-\left[1+\frac{\delta\hat e}{e}-\frac{\delta\hat s_w}{s_w}
+\frac{1}{2}\left(\delta\hat Z_W
+\delta\hat Z_{ii}^{u,L}+\delta\hat Z_{jj}^{d,L}\right)\right]\right.
\nonumber\\
&&{}+\left.
\sum_{a=1}^2\sum_{\sigma=\pm}{\cal M}_a^\sigma
\delta F_a^\sigma\left(M_W,m_{u,i},m_{d,j}\right)\right\},
\label{eq:car}
\end{eqnarray}
where we have exploited the fact that $\delta\hat Z_{ii}^{u,L}$ is real.
We stress that all quantities carrying a caret implicitly depend on the 
specific choice of $i$ and $j$, via $m_{u,i}$ and $m_{d,j}$.
This could be indicated by endowing them with the label $(i,j)$, which we omit
for the time being.
This is important to remember when these quantities are to be summed over.

We then impose the physical OS renormalization condition
\begin{equation}
\tilde{\cal M}_1^{Wu_id_j}=\tilde V_{ij}\hat{\cal M}_1^{Wu_id_j},
\label{eq:ren}
\end{equation}
where $\tilde{\cal M}_1^{Wu_id_j}$ is obtained from ${\cal M}_1^{Wu_id_j}$ of
Eq.~(\ref{eq:amp}) by replacing $V_{ij}$ and $\delta V_{ij}$ with
$\tilde V_{ij}$ and $\delta\tilde V_{ij}$, respectively.
Notice that, at one loop, the renormalization constants in Eq.~(\ref{eq:amp})
are not yet affected by these substitutions.
The salient point is that $\hat{\cal M}_1^{Wu_id_j}$ is UV finite and gauge
independent, since it represents the OS-renormalized $T$-matrix element of a 
physical process.
If we require $\tilde V_{ij}$ to be also UV finite and gauge independent, as
we do, then Eq.~(\ref{eq:ren}) provides an implicit definition of
$\delta\tilde V_{ij}$ with the desired properties.
In the present case, it is particularly simple to solve Eq.~(\ref{eq:ren}) for
$\delta\tilde V_{ij}$, since the form factors $\delta F_a^\sigma$ do not
depend on CKM matrix elements.
We have
\begin{equation}
\frac{\delta\tilde V_{ij}}{\tilde V_{ij}}
=C+\frac{1}{2}\left(\delta\hat Z_{ii}^{u,L}+\delta\hat Z_{jj}^{d,L}\right)
-\frac{1}{2\tilde V_{ij}}
\sum_k\left(\delta Z_{ik}^{u,L\dagger}\tilde V_{kj}
+\tilde V_{ik}\delta Z_{kj}^{d,L}\right),
\label{eq:til}
\end{equation}
where
\begin{equation}
C=\frac{\delta\hat e-\delta e}{e}-\frac{\delta\hat s_w-\delta s_w}{s_w}
+\frac{1}{2}\left(\delta\hat Z_W-\delta Z_W\right).
\label{eq:c}
\end{equation}
In fact, Eq.~(\ref{eq:til}) contains the correct UV divergences and is gauge
independent, as we checked by explicit calculation.
An appealing feature of Eq.~(\ref{eq:til}) is that it only depends on 
self-energies, while the vertex corrections, which are specific for the
considered process, have cancelled.
In this formulation of OS renormalization scheme, all one-loop renormalization
constants are thus expressed in terms of self-energies.
We note in passing that we recover Eq.~(25) of Ref.~\cite{bar} by putting
$C=0$ in Eq.~(\ref{eq:til}).
In other words, the shifts in the renormalization constants of the parameters
and the $W$-boson wave function are not taken into account in Ref.~\cite{bar}.
However, at one loop, we have $\delta\hat e-\delta e=0$, while
$\delta\hat s_w-\delta s_w$ and $\delta\hat Z_W-\delta Z_W$ are by themselves
UV finite and gauge independent, so that the same is true for $C$.
At one loop, it is, therefore, legitimate to put $C=0$ in Eq.~(\ref{eq:til}).
The price to pay is that the physical OS renormalization condition of 
Eq.~(\ref{eq:ren}) must be surrendered.
Although, at first sight, Eq.~(\ref{eq:til}) looks rather complicated, it is
easy to implement in practice, the result being just the right-hand side of
Eq.~(\ref{eq:ren}).

Unfortunately, Eq.~(\ref{eq:til}) and its simplified version of
Ref.~\cite{bar} violate the unitarity condition of Eq.~(\ref{eq:uni}), which
we checked by explicit calculation.
This may also be seen by writing Eq.~(\ref{eq:til}) in the form \cite{den}
\begin{equation}
\delta\tilde V_{ij}
=\sum_k\left(U_{ik}^\dagger\tilde V_{kj}+\tilde V_{ik}D_{kj}\right),
\label{eq:uvd}
\end{equation}
where
\begin{eqnarray}
U_{ik}^\dagger&=&\frac{1}{2}\left[\left(C
+\delta\hat Z_{ii}^{u,L}\right)\delta_{ik}-\delta Z_{ik}^{u,L\dagger}\right],
\nonumber\\
D_{kj}&=&\frac{1}{2}\left[\left(C+\delta\hat Z_{jj}^{d,L}\right)\delta_{kj}
-\delta Z_{kj}^{d,L}\right].
\end{eqnarray}
The unitarity of $\tilde V_{ij}$ would be guaranteed if $U_{ik}^\dagger$ and
$D_{kj}$ were antihermitian matrices.
However, they are not, which already follows from the observation that their
diagonal elements are not purely imaginary, even if the real quantity $C$ is
nullified.

We obtain our final expression for $\delta V_{ij}$ by shifting
$\delta\tilde V_{ij}$ as
\begin{equation}
\delta V_{ij}=\frac{1}{2}\left(\delta\tilde V_{ij}
-\sum_{k,l}V_{ik}\delta\tilde V_{kl}^\dagger V_{lj}\right),
\label{eq:end}
\end{equation}
which has the same UV divergences as $\delta\tilde V_{ij}$, is gauge 
independent, and exactly satisfies Eq.~(\ref{eq:uni}).
Notice that, if $\delta\tilde V_{ij}$ could be represented in the form of 
Eq.~(\ref{eq:uvd}) with antihermitian matrices $U^\dagger$ and $D$, the 
right-hand side of Eq.~(\ref{eq:end}) would be equal to $\delta\tilde V_{ij}$.
Since such a representation is, in fact, possible for the UV divergences of
$\delta\tilde V_{ij}$, this explains why $\delta V_{ij}$ has the same UV
divergences as $\delta\tilde V_{ij}$.
We remark that Eq.~(\ref{eq:end}) represents the infinitesimal form of the
polar decomposition $\tilde V=V\left|\tilde V\right|$, where
$\left|\tilde V\right|=\left(\tilde V^\dagger\tilde V\right)^{1/2}$
\cite{ree}.
In fact, inserting $\tilde V=V^0-\delta\tilde V$ and $V=V^0-\delta V$ into
$V=\tilde V\left(\tilde V^\dagger\tilde V\right)^{-1/2}$ and neglecting terms
beyond one loop, we recover Eq.~(\ref{eq:end}).

Inserting Eq.~(\ref{eq:til}) into Eq.~(\ref{eq:end}) and observing that
$V_{ij}=\tilde V_{ij}=V_{ij}^0$ at the tree level, we obtain
\begin{equation}
\delta V_{ij}=V_{ij}A(i,j)-\sum_{k,l}V_{ik}V_{kl}^\dagger V_{lj}A(l,k)
+\frac{1}{4}\sum_k\left[
\left(\delta Z_{ik}^{u,L}-\delta Z_{ik}^{u,L\dagger}\right)V_{kj}
-V_{ik}\left(\delta Z_{kj}^{d,L}-\delta Z_{kj}^{d,L\dagger}\right)\right],
\label{eq:fin}
\end{equation}
where
\begin{equation}
A(i,j)=\frac{1}{2}\left(\frac{\delta\hat e}{e}
-\frac{\delta\hat s_w}{s_w}\right)
+\frac{1}{4}\left(\delta\hat Z_W
+\delta\hat Z_{ii}^{u,L}+\delta\hat Z_{jj}^{d,L}\right).
\label{eq:a}
\end{equation}
Here, we have used the fact that $A(i,j)$ is real.
The third term in Eq.~(\ref{eq:fin}) agrees with Eq.~(3.16) of
Ref.~\cite{den}, which contains the correct UV divergences, but is known to be
gauge dependent \cite{mad,gam,kni,bar}.
The sum of the first two terms in Eq.~(\ref{eq:fin}) is UV finite and cancels
the gauge dependence of the third term.
If the up-type (down-type) quarks were mass degenerate, then $A(i,j)$ would be
independent of $i$ ($j$), so that the first two terms in Eq.~(\ref{eq:fin}) 
would cancel.
Inserting Eq.~(\ref{eq:fin}) into Eq.~(\ref{eq:amp}), we would then be left
with the hermitian parts of $\delta Z_{ik}^{u,L}$ and $\delta Z_{kj}^{d,L}$,
which are regular in the limit where the masses of the up-type or down-type
quarks coincide \cite{pil}.
Consequently, Eq.~(\ref{eq:amp}) is regular in this limit, as it should.
In the case of exact mass degeneracy, one would, of course, avoid the issue of
CKM-matrix renormalization altogether by setting $V_{ij}^0=\delta_{ij}$ in the
bare Lagrangian.

From the discussion below Eq.~(\ref{eq:c}) it follows that, at one loop, we
may simplify Eq.~(\ref{eq:fin}) by omitting the first three terms on the
right-hand side of Eq.~(\ref{eq:a}), at the expense of abandoning
Eq.~(\ref{eq:ren}).
We consider this as {\it ad hoc}.
While the expressions for $\delta V_{ij}$ proposed in Ref.~\cite{gam} involve
quark self-energies evaluated at $q^2=0$, Eq.~(\ref{eq:fin}) is constructed
from ordinary OS renormalization constants \cite{sir} und thus deserves to be 
referred to as a genuine OS counterterm for the CKM matrix.
A similar comment applies to Ref.~\cite{yam}, where the quark self-energies 
are manipulated by means of a procedure similar to the pinch technique, so 
that the resulting quark wave-function renormalization constants differ from 
the conventional OS ones \cite{sir}.

The decay $W^+\to u_i\bar d_j$ is kinematically forbidden if $i=3$.
We could then derive Eq.~(\ref{eq:fin}) by considering the crossed process,
$u_i\to W^+d_j$.
The advantage of our renormalization procedure is that, owing to its
conceptual transparency, it can be extended straightforwardly to extensions of
the SM involving Majorana neutrinos.
This is the topic of Section~\ref{sec:three}.
Furthermore, we believe that it is likely to carry over to higher orders.

We conclude this section by elaborating an alternative physical OS
renormalization condition for the CKM matrix, which was already mentioned in
Ref.~\cite{den}, namely to demand that the loop-corrected $T$-matrix elements
of four selected $W^+\to u_i\bar d_j$ decays coincide with the respective
tree-level expressions.
One would reasonably pick those decay channels whose partial widths are most
precisely measured.
Specifically, one requires for these choices of $(i,j)$ that
\begin{equation}
{\cal M}_1^{Wu_id_j}={\cal M}_0^{Wu_id_j},
\label{eq:den}
\end{equation}
where ${\cal M}_1^{Wu_id_j}$ is given by Eq.~(\ref{eq:amp}),
${\cal M}_0^{Wu_id_j}=-\left(eV_{ij}/\sqrt2s_w\right){\cal M}_1^-$ is the
tree-level result written in terms of renormalized parameters, and it is
understood that only terms proportional to ${\cal M}_1^-$ are retained.
The other standard matrix elements only enter at one loop, so that their form
factors are UV finite and gauge independent by themselves.
From Eq.~(\ref{eq:amp}), one gleans that
\begin{equation}
\frac{\delta V_{ij}}{V_{ij}}=-\frac{\delta e}{e}+\frac{\delta s_w}{s_w}
-\frac{1}{2}\delta Z_W
-\frac{1}{2V_{ij}}\sum_k\left(\delta Z_{ik}^{u,L\dagger}V_{kj}
+V_{ik}\delta Z_{kj}^{d,L}\right)
-\delta F_1^-\left(M_W,m_{u,i},m_{d,j}\right).
\label{eq:phy}
\end{equation}
By construction, Eq.~(\ref{eq:phy}) contains the correct UV divergences and is
gauge independent.
The residual five counterterms $\delta V_{ij}$ must then be fixed so that
Eq.~(\ref{eq:uni}) is fulfilled.
This is conveniently achieved with the aid of the standard parameterization of
the CKM matrix, which utilizes three angles, $\theta_{12}$, $\theta_{23}$, and
$\theta_{13}$, and one phase, $\delta_{13}$ \cite{cha}; see also Eq.~(11.3) of
Ref.~\cite{pdg}.
In terms of the bare parameters
$\left(\alpha_1^0,\alpha_2^0,\alpha_3^0,\alpha_4^0\right)
\equiv\left(\theta_{12}^0,\theta_{23}^0,\theta_{13}^0,\delta_{13}^0\right)$,
we can write
$V_{ij}^0=f_{ij}(\mbox{\boldmath$\alpha^0$})$.
Substituting $\alpha_k^0=\alpha_k+\delta\alpha_k$, we can identify
$V_{ij}=f_{ij}(\mbox{\boldmath$\alpha$})$, so that
\begin{equation}
\delta V_{ij}=\sum_{k=1}^4\delta\alpha_k
\frac{\partial f_{ij}(\mbox{\boldmath$\alpha$})}{\partial\alpha_k},
\label{eq:par}
\end{equation}
up to higher-order terms.
Equating Eqs.~(\ref{eq:phy}) and (\ref{eq:par}) for the four selected pairs
$(i,j)$, we obtain a linear system of equations, which we can solve for
$\delta\alpha_k$.
The solutions are gauge independent.
The counterterms $\delta V_{ij}$ for the residual pairs $(i,j)$ are then given
by Eq.~(\ref{eq:par}).
They are gauge independent, too.
We checked by explicit calculation that they also contain the correct UV 
divergences.
In conclusion, this alternative OS renormalization condition is physical and
satisfies all three criteria.

An alternative formulation that retains all form factors is obtained by taking
the absolute square and summing over the polarization of the $W^+$ boson and
the spins of the $u_i$ and $\bar d_j$ quarks on both sides of
Eq.~(\ref{eq:den}).
The resulting expression for $\delta V_{ij}/V_{ij}$ differs from
Eq.~(\ref{eq:phy}) by the additional term
$-\sum_{(a,\sigma)}\left(G_a^\sigma/G_1^-\right)
\delta F_a^\sigma\left(M_W,m_{u,i},m_{d,j}\right)$,
where it is summed over $(a,\sigma)=(1,+),(2,+),(2,-)$ and
$G_a^\sigma=\sum_{\rm pol}{\cal M}_1^{-\dagger}{\cal M}_a^{\sigma}$ are real
functions of $M_W$, $m_{u,i}$, and $m_{d,j}$, which may be found in Eq.~(2.3)
of Ref.~\cite{kni}.
This term is UV finite and gauge independent.
The determination of the residual counterterms $\delta V_{ij}$ then proceeds
as explained above.

Obvious drawbacks of Eq.~(\ref{eq:den}) and its variant discussed in the 
preceding paragraph are that they destroy the symmetry of Eq.~(\ref{eq:amp})
with respect to the quark families \cite{den} and that the resulting
expressions for $\delta V_{ij}$ involve vertex corrections, which are specific
for the selected processes, whereas all other renormalization constants of the
SM can be expressed in terms of self-energies.
While Eq.~(\ref{eq:ren}) does not suffer from these drawbacks, it entails the
minor complication that one needs to consider a reference theory with zero
mixing.
Notwithstanding, we find the OS renormalization prescription presented in the
first part of this section preferable.

\section{Renormalization of the lepton mixing matrices in Majorana-neutrino 
theories}\label{sec:three}

In this section, we consider a minimal, renormalizable extension of the SM,
based on the SU(2)${}_I\otimes{}$U(1)${}_Y$ gauge group, that can naturally
accommodate heavy Majorana neutrinos \cite{apo,pil} and propose physical OS
renormalization conditions for its lepton mixing matrices that satisfy all
three criteria enumerated in Section~\ref{sec:one}.
To this end, we need to generalize the formalism proposed in 
Section~\ref{sec:two}.
As a new feature, we encounter the constraint that the texture zero in the
see-saw mass matrix should be preserved by the renormalization procedure in
order not to increase the number of independent parameters.

To start with, we summarize the basic features of the lepton sector of the
Majorana-neutrino theory under consideration, adopting the notation from
Refs.~\cite{apo,pil}.
For generality, we allow for an arbitrary number $N_G$ of fermion generations.
Similarly to the SM, each lepton family contains one weak-isospin ($I$)
doublet $(\nu_{L,i}^\prime,l_{L,i}^\prime)$ of left-handed states with weak 
hypercharge $Y=-1$ and one right-handed charged-lepton state $l_{R,i}^\prime$
with $I=0$ and $Y=-2$.
In addition, there is a total of $N_R$ right-handed neutrinos
$\nu_{R,i}^\prime$ with $I=Y=0$.
Here, the primes are to remind us that we are dealing with weak-interaction
eigenstates.
Deviating from Refs.~\cite{apo,pil}, we require that $N_R$ is a multiple of
$N_G$ and that there are $N_R/N_G$ right-handed neutrinos in each lepton
family, so that all lepton families have the same structure.
The quark families are taken to be of the SM type.
The bare Lagrangian contains the neutrino mass terms
\begin{equation}
{\cal L}_Y^{0,\nu}
=-\frac{1}{2}\left(\bar\nu_L^{\prime0},\bar\nu_R^{\prime0C}\right)
M^{\prime0,\nu}\left(
\begin{array}{c}
\nu_L^{\prime0C} \\
\nu_R^{\prime0}
\end{array}
\right)+\mbox{h.c.},
\end{equation}
where
$\nu_L^{\prime0}=\left(\nu_{L,1}^{\prime0},\ldots,\nu_{L,N_G}^{\prime0}
\right)^T$,
$\nu_R^{\prime0}=\left(\nu_{R,1}^{\prime0},\ldots,\nu_{R,N_R}^{\prime0}
\right)^T$,
the superscript $C$ denotes charge conjugation, and $M^{\prime0,\nu}$ is a
complex, symmetric mass matrix of the from
\begin{equation}
M^{\prime0,\nu}=\left(
\begin{array}{cc}
m_L^{\prime0} & m_D^{\prime0} \\
m_D^{\prime0T} & m_M^{\prime0}
\end{array}
\right).
\end{equation}
Unless the SM Higgs sector is supplemented by an additional weak-isospin
triplet of Higgs fields, gauge invariance enforces $m_L^{\prime0}=0$
\cite{sch}.
In the following, we assume that $m_L^{\prime0}=0$.
This allows for the see-saw mechanism \cite{yan} to operate.
The neutrino mass matrix $M^{\prime0,\nu}$ can always be diagonalized through
a unitary transformation $U^{0,\nu}$ as
\begin{equation}
U^{0,\nu T}M^{\prime0,\nu}U^{0,\nu}
={\rm diag}\left(m_{n,1}^0,\ldots,m_{n,N_G+N_R}^0\right),
\end{equation}
where $m_{n,i}^0\ge0$ at the tree level.
The corresponding mass eigenstates are given by 
\begin{eqnarray}
\left(
\begin{array}{c}
\nu_L^{\prime0} \\
\nu_R^{\prime0C}
\end{array}
\right)_i&=&
\sum_{j=1}^{N_G+N_R}U^{0,\nu\ast}_{ij}n_{L,j}^0,
\nonumber\\
\left(
\begin{array}{c}
\nu_L^{\prime0C} \\
\nu_R^{\prime0}
\end{array}
\right)_i&=&
\sum_{j=1}^{N_G+N_R}U^{0,\nu}_{ij}n_{R,j}^0.
\end{eqnarray}
Here, the first $N_G$ mass eigenstates, $\nu_i\equiv n_i$ ($i=1,\dots,N_G$),
are identified with the ordinary light neutrinos (assuming that $N_G=3$), and
the remaining $N_R$ states, $N_i\equiv n_{N_G+i}$ ($i=1,\dots,N_R$), represent
the new neutral leptons predicted by the theory.
The latter are the heavy Majorana neutrinos, which have not yet been 
discovered.
The diagonalization of the charged-lepton mass matrix proceeds as in the quark
case discussed in Section~\ref{sec:two}.

In this Majorana-neutrino theory, mixing effects enter via the interactions of
the charged leptons $l_i$ and Majorana neutrinos $n_i$ with the $W$, $Z$, and
Higgs bosons.
The bare Lagrangian of the charged-current interaction involves the
$N_G\times(N_G+N_R)$ mixing matrix $B^0$, while the ones of the
neutral-current and Yukawa interactions involve the $(N_G+N_R)\times(N_G+N_R)$
mixing matrix $C^0$.
Specifically, we have
\begin{eqnarray}
B_{ij}^0&=&\sum_{k=1}^{N_G}V_{ik}^{0,l}U_{kj}^{0,\nu\ast},
\nonumber\\
C_{ij}^0&=&\sum_{k=1}^{N_G}U_{ik}^{0,\nu T}U_{kj}^{0,\nu\ast},
\label{eq:bc}
\end{eqnarray}
where $V^{0,l}$ is the unitary $N_G\times N_G$ matrix relating the
weak-interaction and mass eigenstates of the bare left-handed charged-lepton
fields,
\begin{equation}
l_{L,i}^0=\sum_{j=1}^{N_G}V_{ij}^{0,l}l_{L,j}^{\prime0}.
\end{equation}
Notice that the summations in Eq.~(\ref{eq:bc}) stop at $k=N_G$, rather than 
at $k=N_G+N_R$, thereby projecting the neutrino state vector onto its 
non-isosinglet components.
From Eq.~(\ref{eq:bc}), it follows that $B^0$ is pseudo-unitary, in the sense 
that it satisfies the relationships
\begin{eqnarray}
\sum_{k=1}^{N_G+N_R}B_{ik}^0B_{kj}^{0\dagger}&=&\delta_{ij},
\label{eq:bbd}\\
\sum_{k=1}^{N_G}B_{ik}^{0\dagger}B_{kj}^0&=&C_{ij}^0.
\label{eq:bdb}
\end{eqnarray}
From Eq.~(\ref{eq:bc}), it also follows that
\begin{eqnarray}
C_{ij}^{0\dagger}=\sum_{k=1}^{N_G+N_R}C_{ik}^0C_{kj}^0&=&C_{ij}^0,
\nonumber\\
\sum_{k=1}^{N_G+N_G}B_{ik}^0C_{kj}^0&=&B_{ij}^0.
\label{eq:cor}
\end{eqnarray}
Notice that Eq.~(\ref{eq:cor}) can also be derived from Eqs.~(\ref{eq:bbd})
and (\ref{eq:bdb}), reflecting the fact that $C^0$ is but an auxiliary
quantity, which is completely fixed once $B^0$ is.
In the see-saw scenario, with $m_L^{\prime0}=0$, there are additional
relationships between $B^0$, $C^0$, and $m_{n,i}^0$, namely
\begin{eqnarray}
\sum_{k=1}^{N_G+N_R}B_{ik}^0m_{n,k}^0B_{kj}^{0T}&=&0,
\label{eq:bmb}\\
\sum_{k=1}^{N_G+N_R}B_{ik}^0m_{n,k}^0C_{kj}^{0T}&=&0,
\label{eq:bmc}\\
\sum_{k=1}^{N_G+N_R}C_{ik}^0m_{n,k}^0C_{kj}^{0T}&=&0.
\label{eq:cmc}
\end{eqnarray}
Notice that Eqs.~(\ref{eq:bmc}) and (\ref{eq:cmc}) follow from 
Eq.~(\ref{eq:bmb}) with the aid of Eq.~(\ref{eq:bdb}).
For later use, we remark here that Eqs.~(\ref{eq:bmb})--(\ref{eq:cmc}) are
valid for arbitrary values of $m_{n,i}^0$.

We now turn to the renormalization of the lepton mixing matrices and write
$B_{ij}^0=B_{ij}+\delta B_{ij}$, $C_{ij}^0=C_{ij}+\delta C_{ij}$, and
$m_{n,i}^0=m_{n,i}+\delta m_{n,i}$.
The third criterion implies that the renormalized mixing matrices $B$ and $C$
must satisfy relations analogous to Eqs.~(\ref{eq:bbd}) and (\ref{eq:bdb}).
Then, they automatically satisfy relations analogous to Eq.~(\ref{eq:cor}).
This leads to the following conditions:
\begin{eqnarray}
\sum_{k=1}^{N_G+N_R}\left(\delta B_{ik}B_{kj}^\dagger
+B_{ik}\delta B_{kj}^\dagger\right)&=&0,
\label{eq:bun}\\
\sum_{k=1}^{N_G}\left(\delta B_{ik}^\dagger B_{kj}
+B_{ik}^\dagger\delta B_{kj}\right)&=&\delta C_{ij},
\label{eq:cun}
\end{eqnarray}
up to higher-order terms.
Equation~(\ref{eq:bun}) is the analogue of Eq.~(\ref{eq:uni}) for the CKM
matrix, while Eq.~(\ref{eq:cun}) tells us that $\delta C_{ij}$ is fixed once
$\delta B_{ij}$ is.
Having found an expression for $\delta B_{ij}$ that contains the correct UV 
divergences and satisfies Eq.~(\ref{eq:bun}), we are in general faced with the
situation that radiative corrections destroy the texture zero in the
see-saw mass matrix.
It is, therefore, necessary to impose the additional condition that $B$, $C$,
and $m_{n,i}$ must satisfy relations analogous to
Eqs.~(\ref{eq:bmb})--(\ref{eq:cmc}), in order not to increase the number of
independent parameters.
As in bare case, it is sufficient to require the analogue of
Eq.~(\ref{eq:bmb}), which is satisfied if
\begin{equation}
\sum_{k=1}^{N_G+N_R}\left(\delta B_{ik}m_{n,k}B_{kj}^T
+B_{ik}\delta m_{n,k}B_{kj}^T+B_{ik}m_{n,k}\delta B_{kj}^T\right)=0,
\label{eq:dbmb}
\end{equation}
up to higher-order terms.
In Ref.~\cite{pil}, this additional condition was not imposed, with the 
consequence that the second criterion was not fulfilled.
This would in general also happen in possible generalizations of the
renormalization prescriptions for the CKM matrix recently proposed in
Refs.~\cite{mad,gam,bar} if this additional condition were not implemented.

In the following, we present a physical OS renormalization prescription for
the lepton mixing matrices that satisfies all three criteria and, in 
particular, guarantees that $m_L^{\prime}=0$ if $m_L^{0\prime}=0$ is chosen.
We start by observing that in the Majorana-neutrino theory under
consideration, with $m_L^{0\prime}=0$, the parameters $B_{ij}^0$ are in
general functions of $m_{n,k}^0$ and some bare angles $\theta_l^0$ and phases
$\delta_m^0$ fixing the remaining degrees of freedom, so that
Eqs.~(\ref{eq:bbd}) and (\ref{eq:bmb}) are fulfilled for all values of
$m_{n,k}^0$, $\theta_l^0$, and $\delta_m^0$.
That is, we have some parameterization
$B_{ij}^0=f_{ij}\left(\mbox{\boldmath$m_n^0$},\mbox{\boldmath$\alpha^0$}
\right)$,
where
$\mbox{\boldmath$\alpha^0$}=(\mbox{\boldmath$\theta^0$},
\mbox{\boldmath$\delta^0$})$.
In the most general case, there are $2N_G(N_R-1)$ independent parameters
$\alpha_l^0$.
On the other hand, the quark mixing matrix of this theory comprises
$(N_G-1)^2$ independent parameters.
Henceforth, we confine ourselves to parameterizations of $B^0$ with the 
property that
$\hat B_{ij}^0=f_{ij}\left(\mbox{\boldmath$m_n^0$},\mbox{\boldmath$0$}
\right)\equiv\hat f_{ij}\left(\mbox{\boldmath$m_n^0$}\right)$
does not mix different lepton families.
Although this requirement restricts the choice of parameterizations, we
believe that all phenomenologically interesting scenarios can still be
described.
Notice that $\hat B^0$ does not imply zero mixing, as this would in general
violate Eq.~(\ref{eq:bmb}).
This rather means that, depending on the particular choice of
parameterization, $\hat B^0$ is as close as possible to the zero-mixing case,
while it still satisfies Eq.~(\ref{eq:bmb}).
The only parameters that the bare Lagrangian of the Majorana-neutrino theory
with $\hat B^0$ contains in addition to those of the SM are $m_{n,k}^0$.
This theory can, therefore, be renormalized according to the well-established
OS renormalization scheme \cite{sir}, without introducing any other additional
counterterms than $\delta m_{n,k}$ \cite{pil}.
Similarly to
$\left(s_w^0\right)^2=1-\left(M_W^0\right)^2/\left(M_Z^0\right)^2$ in the SM,
the parameters $\hat B_{ij}^0=\hat f_{ij}\left(\mbox{\boldmath$m_n^0$}\right)$
are then not treated as basic ones, but rather as convenient abbreviations.
Writing $\hat B_{ij}^0=\hat B_{ij}+\delta\hat B_{ij}$, we can identify
$\hat B_{ij}=\hat f_{ij}\left(\mbox{\boldmath$m_n$}\right)$, so that
\begin{equation}
\delta\hat B_{ij}=\sum_{k=1}^{N_G+N_R}\delta m_{n,k}
\frac{\partial\hat f_{ij}(\mbox{\boldmath$m_n$})}{\partial m_{n,k}},
\end{equation}
up to higher-order terms.
In this way, $\hat B_{ij}$ is automatically UV finite, gauge independent, and
pseudo-unitary.
Moreover, it satisfies Eq.~(\ref{eq:bmb}) written with $\hat B_{ij}$ and 
$m_{n,k}$.
The counterterms $\delta\hat B_{ij}$ cancel the additional UV divergences that
arise through the intrafamily lepton mixing induced by $\hat B^0$.

Next, we return to the theory with $B^0$ by reintroducing the parameters
$\alpha_k^0$.
Similarly to Section~\ref{sec:two}, we proceed in two steps to fix
$\delta B_{ij}$.
We first construct intermediate counterterms $\delta\tilde B_{ij}$ that
contain the correct UV divergences and are gauge independent.
To this end, we introduce a physical OS renormalization condition relating the
loop-corrected results for the partial width of the decay
$W^-\to l_i^-n_j$ (or $n_j\to W^+l_i^-$) calculated in the theories with $B^0$
and $\hat B^0$.
The resulting expressions for $\delta\tilde B_{ij}$ and
$\tilde B_{ij}=B_{ij}^0-\delta\tilde B_{ij}$ will in general not satisfy the
pseudo-unitarity condition of Eq.~(\ref{eq:bun}).
Then, we obtain our final expression for $\delta B_{ij}$ by adjusting
$\delta\tilde B_{ij}$ by UV-finite, gauge-independent terms so that
Eq.~(\ref{eq:bun}) is fulfilled.

In the general theory with $B^0$, the one-loop-corrected $T$-matrix element of
the decay $W^-\to l_i^-n_j$ has the form
\begin{eqnarray}
{\cal M}_1^{Wl_in_j}&=&-\frac{eB_{ij}}{\sqrt2s_w}\left\{
{\cal M}_1^-\left[1+\frac{\delta e}{e}-\frac{\delta s_w}{s_w}
+\frac{\delta B_{ij}}{B_{ij}}+\frac{1}{2}\delta Z_W\right.\right.
\nonumber\\
&&{}+\left.\left.
\frac{1}{2B_{ij}}\sum_k\left(\delta Z_{ik}^{l,L\dagger}B_{kj}
+B_{ik}\delta Z_{kj}^{n,L}\right)\right]
+\sum_{a=1}^2\sum_{\sigma=\pm}{\cal M}_a^\sigma
\delta F_a^\sigma\right\},
\label{eq:wln}
\end{eqnarray}
where the form factors $\delta F_a^\sigma$ are now functions of $M_W$, $M_Z$,
$M_H$, $m_{l,k}$, $m_{n,l}$, $B_{mn}$.
In the minimal-mixing theory with $\hat B^0$, the decay $W^-\to l_i^-n_j$ is
only allowed if $l_i$ and $n_j$ belong to the same lepton family.
If this is not the case, then we need to redefine this theory by interchanging
$l_i$ and $l_j$.
We are entitled to do so because, in the general-mixing theory with $B^0$, the
assignment of mass eigenstates to fermion generations is, as a matter of
principle, arbitrary.
The expression for $\hat{\cal M}_1^{Wl_in_j}$ in the reference theory thus 
defined emerges from Eq.~(\ref{eq:wln}) by substituting $B_{ij}$ with
$\hat B_{ij}$ and placing a caret on each renormalization constant.
The form factors, which we denote by $\delta\hat F_a^\sigma$, are now 
independent of the masses $m_{l,k}$ and $m_{n,l}$ of those leptons $l_k$ and
$n_l$ that do not belong the same family as $l_i$ and $n_j$.
In contrast to $\hat{\cal M}_1^{Wu_id_j}$ of Eq.~(\ref{eq:car}),
$\hat{\cal M}_1^{Wl_in_j}$ contains the overall factor $\hat B_{ij}$, which
in general differs from unity. 
We thus need to generalize the physical OS renormalization condition of
Eq.~(\ref{eq:ren}) to read
\begin{equation}
\tilde{\cal M}_1^{Wl_in_j}=\frac{\tilde B_{ij}}{\hat B_{ij}}
\hat{\cal M}_1^{Wl_in_j},
\label{eq:maj}
\end{equation}
where it is understood that we only retain terms proportional to
${\cal M}_1^-$.
The other standard matrix elements only enter at one loop, so that their form
factors are UV finite and gauge independent by themselves.
Below, we outline an alternative approach that also includes these form
factors.
We note that $\hat B_{ij}\ne0$ because the reference theory is arranged so 
that $\hat B_{ij}^0\ne0$.
Solving Eq.~(\ref{eq:maj}) for $\delta\tilde B_{ij}$, we find
\begin{eqnarray}
\frac{\delta\tilde B_{ij}}{\tilde B_{ij}}
&=&C+\frac{\delta\hat B_{ij}}{\hat B_{ij}}
+\frac{1}{2\hat B_{ij}}\sum_k\left(\delta\hat Z_{ik}^{l,L\dagger}\hat B_{kj}
+\hat B_{ik}\delta\hat Z_{kj}^{n,L}\right)
-\frac{1}{2\tilde B_{ij}}
\sum_k\left(\delta Z_{ik}^{l,L\dagger}\tilde B_{kj}
+\tilde B_{ik}\delta Z_{kj}^{n,L}\right)
\nonumber\\
&&{}+\delta\hat F_1^--\delta F_1^-,
\label{eq:bti}
\end{eqnarray}
where $C$ is defined in Eq.~(\ref{eq:c}).
By construction, Eq.~(\ref{eq:bti}) contains the correct UV divergences and is
gauge independent, as we also checked by explicit calculation for
representative choices of $B^0$.
In contrast to the SM, where all OS renormalization constants can be expressed
in terms of self-energies \cite{sir}, Eq.~(\ref{eq:bti}) involves also vertex
corrections.
Similarly to Eq.~(\ref{eq:til}), at one loop, we can simplify
Eq.~(\ref{eq:bti}) by discarding $C$, at the expense of sacrificing
Eq.~(\ref{eq:maj}).

An alternative formulation that retains all form factors is obtained by taking
the absolute square and summing over the polarization of the $W^-$ boson and
the spins of the $l_i^-$ and $n_j$ leptons on both sides of
Eq.~(\ref{eq:maj}).
The resulting expression for $\delta\tilde B_{ij}/\tilde B_{ij}$ differs from
Eq.~(\ref{eq:bti}) by the additional term
$\sum_{(a,\sigma)}\left(G_a^\sigma/G_1^-\right)
\left(\delta\hat F_a^\sigma-\delta F_a^\sigma\right)$,
where it is summed over $(a,\sigma)=(1,+),(2,+),(2,-)$.
Since this term is UV finite and gauge independent, we have the option of
discarding it along with $C$, with the same consequence.

Unfortunately, Eq.~(\ref{eq:bti}) and its variants discussed above violate the
pseudo-unitarity condition of Eq.~(\ref{eq:bun}).
Similarly to Eq.~(\ref{eq:end}), this problem can be fixed by the redefinition
\begin{equation}
\delta B_{ij}=\frac{1}{2}\left(\delta\tilde B_{ij}
-\sum_{k,l}B_{ik}\delta\tilde B_{kl}^\dagger B_{lj}\right),
\label{eq:db}
\end{equation}
which has the same UV divergences as $\delta\tilde B_{ij}$, is gauge 
independent, and exactly satisfies Eq.~(\ref{eq:bun}).
Inserting Eq.~(\ref{eq:bti}) into Eq.~(\ref{eq:db}), we obtain an expression
for $\delta B_{ij}$ that extends Eq.~(6.1) of Ref.~\cite{pil} by UV-finite,
gauge-dependent terms, which compensate the gauge dependence of that equation.
Our final result for $B_{ij}$ originates from a genuine OS renormalization
condition and satisfies all three criteria.

We conclude this section by pointing out that the alternative physical OS
renormalization condition for the CKM matrix worked out at the end of
Section~\ref{sec:two} can straightforwardly be extended to the
Majorana-neutrino case.
To this end, one chooses a parameterization
$B_{ij}^0=f_{ij}\left(\mbox{\boldmath$m_n^0$},\mbox{\boldmath$\alpha^0$}
\right)$
that satisfies Eqs.~(\ref{eq:bbd}) and (\ref{eq:bmb}) and selects as many
$W^+\to l_i^-n_j$ decay channels as there are parameters $\alpha_k^0$, namely
$2N_G(N_R-1)$.
For these processes, one then defines $\delta B_{ij}$ by nullifying the
loop corrections proportional to ${\cal M}_1^-$ in Eq.~(\ref{eq:wln}).
As in the quark case, one can alternatively equate the absolute squares of the
tree-level and loop-corrected $T$-matrix elements after summation over
polarization and spins.
The residual expressions for $\delta B_{ij}$ are then determined according to
the procedure outlined at the end of Section~\ref{sec:two}, except that we now
have, up to higher-order terms,
\begin{equation}
\delta B_{ij}
=\sum_{k=1}^{N_G+N_R}\delta m_{n,k}\frac{\partial
f_{ij}(\mbox{\boldmath$m_n$},\mbox{\boldmath$\alpha$})}{\partial m_{n,k}}
+\sum_{k=1}^{2N_G(N_R-1)}\delta\alpha_k\frac{\partial
f_{ij}(\mbox{\boldmath$m_n$},\mbox{\boldmath$\alpha$})}{\partial\alpha_k},
\label{eq:dbp}
\end{equation}
which also involves the known expressions for $\delta m_{n,k}$ \cite{pil}.
As in the quark case, this alternative OS renormalization prescription is
physical and satisfies all three criteria.
At the same time, it guarantees that Eq.~(\ref{eq:dbmb}) is fulfilled.
However, similarly to the quark case, it breaks the symmetry between the
lepton families.
For this reason, we advocate the OS renormalization prescription presented in
the first part of this section.

\section{Conclusions}\label{sec:four}

We proposed physical OS renormalization conditions for the CKM matrix of the 
SM and for the lepton mixing matrices of a favourable class of see-saw-type
Majorana-neutrino theories, in which all fermion generations have the same
structure.
Apart from being UV finite, the resulting renormalized mixing matrices are
gauge independent and (pseudo)unitary.

In the SM, our strategy was to select a suitable charged-current process, such
as the decay $W^+\to u_i\bar d_j$, and to match the loop-corrected $T$-matrix
element of the full theory with the one of a reference theory, in which the
$u_i$ and $d_j$ quarks are arranged to belong to the same fermion generation
and quark mixing is switched off.
Since the latter is UV finite and gauge independent, this equality defines a
counterterm for the CKM matrix that has the correct UV divergences and is
gauge independent.
However, the corresponding renormalized CKM matrix is not unitary.
In a second step, unitarity is installed by a UV-finite, gauge-independent
shift of the counterterm.

This procedure was then generalized to see-saw-type Majorana-neutrino
theories.
Here, one faces the complication that the lepton mixing matrices do not only
depend on mixing angles and phases, which are responsible for interfamily
mixing, but also on neutrino masses.
Thus, nullifying the mixing angles and phases leads us to a reference theory
with intrafamily mixing, rather than zero mixing.
The lepton mixing matrices of this reference theory can be renormalized by
shifting the bare neutrino masses according to the usual OS renormalization
scheme.
In this way, the texture zero in the neutrino mass matrix is preserved, and so
is gauge invariance.
The matching and unitarization procedures can then be carried out in analogy 
to the SM case.
In general Majorana-neutrino theories without see-saw mechanism, where the
texture zero is traded against an additional Higgs triplet, the lepton mixing
matrices can be taken to be independent of neutrino masses, which considerably
simplifies the renormalization procedure.

We explicitly worked at one loop, but we believe that our formalism is likely
to carry over to higher orders.
From the phenomenological point of view, a two-loop analysis of the CKM matrix
does not appear to be necessitated by the experimental precision to be
achieved in the forseeable future.
In the Majorana-neutrino case, such a two-loop analysis lacks motivation
before the neutrino puzzle is solved and the underlying pattern of the
neutrino sector is fully understood.

\vspace{1cm}
\noindent
{\bf Acknowledgements}
\smallskip

We thank A. Denner, W. Hollik, and E. Kraus for very useful discussions.
The work of K.-P.O.D. was supported by the Max-Planck-Institut f\"ur Physik
through a PhD fellowship.
The work of B.A.K. was supported in part by the Deutsche
Forschungsgemeinschaft through Grant No.\ KN~365/1-1, by the
Bundesministerium f\"ur Bildung und Forschung through Grant No.\ 05~HT1GUA~4,
and by the European Commission through the Research Training Network
{\it Quantum Chromodynamics and the Deep Structure of Elementary Particles}
under Contract No.\ ERBFMRX-CT98-0194.

\end{document}